# Technical Report:

# Evaluation of Machine Learning Fameworks on Finis Terrae II

**Abstract:** Machine Learning (ML) and Deep Learning (DL) are two technologies used to extract representations of the data for a specific purpose. ML algorithms take a set of data as input to generate one or several predictions. To define the final version of one model, usually there is an initial step devoted to train the algorithm (get the right final values of the parameters of the model). There are several techniques, from supervised learning to reinforcement learning, which have different requirements. On the market, there are some frameworks or APIs that reduce the effort for designing a new ML model. In this report, using the benchmark DLBENCH, we will analyse the performance and the execution modes of some well-known ML frameworks on the Finis Terrae II supercomputer when supervised learning is used. The report will show that placement of data and allocated hardware can have a large influence on the final time-to-solution.

| | |
|---|---|
| Document Id.: | CESGA-2017-004 |
| Date: | **December 27th, 2017** |
| Task: | |
| Responsible: | **Andrés Gómez Tato** |
| Status: | **Final** |

# Evaluation of Machine Learning Frameworks on Finis Terrae II

## Galicia Supercomputing Centre

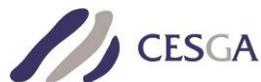

# Index

# List of Figures

# List of Tables

# 1 Introduction

Machine Learning is a set of techniques used to extract models, representations or predictions from data. These techniques include a large variety of algorithms, from the least-square method proposed by the French mathematician Adrien-Marie Legendre in 1805 to the last techniques based on several layers of mathematical operations known as Deep Learning. A review of these techniques is out of the scope of this report, so the interested reader should use other references to learn about.

There are several techniques to design and parameterise Machine Learning algorithms, known technically as learning or training. Training can be: supervised (when there are data which are marked with the expected result, called labels); unsupervised (when the training selects automatically the solution, as in some clustering algorithms, where if one case belongs or not to a cluster – or set – is decided by the algorithm); semi-supervised (a mixture of previous methods); or reinforcement learning (when the algorithm can interact with the system to model during the training).

Training is a compute intensive task, specially supervised, because it is based on the mathematical optimisation of a cost function regarding the free parameters of the Machine Learning model, which can be millions. This training requires thousands of steps to achieve a correct value for the free parameters, where the forward pass (make the computations starting from the data to the solution) and the backward (calculating the derivative of the cost function regarding each free parameter) must be executed. Additionally, during the initial design steps, the number of free parameters (or weights) are not known and must be searched, which means that several training jobs must be carried on to find the best model. This technique is known as hyperparametric search, and has been presented in other CESGA Technical Report[1].

Because both hyperparametric search and training of one Machine Learning algorithm is a compute intensive job, using supercomputers as Finis Terrae II is an advantage that allows researchers and engineers to decrease the time to solution or, even, to provide them with enough capacity to process the data, which can be very large. Due to the increased demand, CESGA has installed many of the commonly used Machine Learning packages as Tensorflow, Caffe, CNTK or Theano. However, Finis Terrae II is a hybrid platform which has different architectures (nodes with and without GPUs) and capabilities (i.e., fast storage as Lustre). The availability of different capabilities and software frameworks deserves a careful analysis to select the combination of resources to extract the best performance and time-to-solution. This report summarises the result of an initial benchmark of Machine Learning frameworks which provides an insight about the possibilities of HPC facilities for supervised training.

The report is divided in four sections. The first one describes the used resources. Second section explains the used methodology and continues with a third one dedicated to the main numerical results. Final forth section presents the main conclusions.

---

[1] Ferro, G and Cotelo, C.: Hyper-parametric search using HPC infrastructures for Tensorflow. CESGA-2017-001. https://www.cesga.es/es/biblioteca/downloadAsset/id/797

Certified entity ISO 9001　　　　　　　　　　　　　　　　　　　　　　　　　　　Page 6 of 24

# 2 Resources description

## *2.1 Finis Terrae II*

Finis Terrae II is a hybrid supercomputer that was deployed at CESGA during the first two months of 2017 (was available for the researchers at Feb. 2017). It is composed of several nodes connected by a fast Infiniband FBR in a fat network architecture, which allows the nodes to connect each other with low latency and high bandwidth. Additionally, a LUSTRE filesystem (760TB of capacity, more than 20GB/s of I/O throughput, based on 4 OSS and 480 2TB HDD) is connected to the network and shared among all the nodes. All the nodes can also access to a permanent storage that is connected through Ethernet and mounted using NFS. This large storage is provided by a EMC2 VNX5700 with 2TB NL-SAS disk, and the same cabinet provides limited home capacity to each user, but in this case with faster disk (15K RAM SAS with 600GB capacity)[2].

There are different nodes, which differ in the CPU and additional hardware and is continuously evolving. The full configuration in September and October 2017 (when the benchmark was executed) is shown in the next table:

| Nodes | CPU | RAM | Net | Additional Hardware |
|---|---|---|---|---|
| 298 thin nodes | 2xIntel 2680v3 | 128GB | 2 1GbE<br>1 Infiniband FDR@56Gbps | 1TB disk |
| 4 GPU nodes | 2xIntel 2680v3 | 128GB | 2 1GbE<br>1 Infiniband FDR@56Gbps | 1 TB disk<br>2 NVIDIA© K80 |
| 2 Phi nodes | 2xIntel 2680v3 | 128GB | 2 1GbE<br>1 Infiniband FDR@56Gbps | 1 TB Disk<br>2 Intel Xeon© Phi 7120P |
| 1 FAT node | 8 xs Intel Haswell 8867v3 | 4096GB | 1 Infiniband FDR@56Gbps | 24 SAS disk of 1,2 TB and 2 SAS 300 GB |
| 4 Login/transfer | 2xIntel 2680v3 memoria | 128 GB | 2x10Gbit Ethernet<br>1xInfiniband FDR@56Gbps | 2x1TB disks |
| 4 | 2xIntel 2650v3 | 128GB | 1GbE | 4TB HDD |
| 4 | 2xIntel 2650v3 | 128GB | 1GbE | 4TB HDD<br>1 NVIDIA KGrid |

**Table 1: Summary of Finis Terrae II architecture**

All the nodes have Red Hat Operating System (Red Hat Enterprise Linux Server release 6.7 (Santiago) version), deployed by Bull. The execution of jobs in each node is managed by a Slurm load balancer

---

[2] This storage has been substituted by another new system during December of 2017.

(slurm 14.11.10-Bull.1.0). NVIDIA K80 GPUs are installed with Driver Version 375.26 and Slurm manages each GPU inside a K80 as a separate device, allowing up to four simultaneous jobs in one node or a single job with up to 4 GPUs. GPUs must be explicitly demanded for each job and Slurm allocates them. The topology of the GPU connections is shown in Figure 1.

|      | GPU0 | GPU1 | GPU2 | GPU3 |
|------|------|------|------|------|
| GPU0 |      | PIX  | SOC  | SOC  |
| GPU1 | PIX  |      | SOC  | SOC  |
| GPU2 | SOC  | SOC  |      | PIX  |
| GPU3 | SOC  | SOC  | PIX  |      |

**Figure 1: Topology of the NVIDIA K80 connections inside a Finis Terrae node. PIX means that are connected to the same PCI socket and SOC that the connection must use the CPU.**

Machine Learning frameworks have been compiled with *gcc* version 4.9.1. To manage the environment of each application, Lmod[3] is used.

| Framework | Version | Compiler | Additional Packages |
|---|---|---|---|
| Caffe | 1.0 | Gcc/4.9.1 | boost/1.60.0 |
|   |   |   | hdf5/1.8.16 |
|   |   |   | NVIDIA cuda/8.0 |
|   |   |   | NVIDIA cuDNN/6.0 |
|   |   |   | NVIDIA NCCL/1.3.4 |
|   |   |   | Intel MKL/2017.0 |
|   |   |   | Anaconda2/4.0.0 |
| Caffe-intel | 06.09.2017 | Gcc/4.9.1 | boost/1.60.0 |
|   |   |   | hdf5/1.8.16 |
|   |   |   | Opencv/3.1.0 |
|   |   |   | Anaconda2/4.0.0 |
|   |   |   | Intel MPI/5.1 |
|   |   |   | Intel MKL/2017.0 |
| Tensorflow (CPU-Only) | 1.3.1 | Gcc/4.9.1 | Oracle JDK/1.8.0 |
|   |   |   | openssl/1.0.2f |

---

[3] https://www.tacc.utexas.edu/research-development/tacc-projects/lmod

| Framework | Version | Compiler | Additional Packages |
|---|---|---|---|
| | | | anaconda2/4.0.0 |
| | | | Intel MKL/11.3 |
| Tensorflow | 1.3.1 | Gcc/4.9.1 | Oracle JDK/1.8.0 |
| | | | openssl/1.0.2f |
| | | | anaconda2/4.0.0 |
| | | | Intel MKL/11.3 |
| | | | NVIDIA cuda/8.0 |
| | | | NVIDIA cuDNN/6.0 |

Table 2: Summary of Machine Learning packages available in Finis Terrae II

## 2.2 DLBENCH

DLBENCH (Benchmarking State-of-the-Art Deep Learning Software Tools[4]) [1] is a benchmark developed by the Hong Kong Baptist University which permits an easy management of Machine Learning experiments executing commonly used Neural Networks both fully connected, convolutional or recurrent. It supports directly Caffe[2], Tensorflow[3], CNTK[4], Torch[5] and MXNet[6] and adds the possibility of adapting it to use another frameworks. Table 3 shows the supported networks for each Machine Learning package.

| Net | Type | Number of weights | Caffe | Caffe-intel | Tensorflow | CNTK |
|---|---|---|---|---|---|---|
| FCN5 | Fully Connected | 14,205,962 | Yes | Yes | Yes | Yes |
| ResNet-56 | Convolutional | 853,018 | Yes | Yes | Yes | Yes |
| AlextNet-R | Convolutional | 85,098 | Yes | Yes | Yes | Yes |
| LSTM | Recurrent | 5,655,312 | No | No | Yes | Yes |

Table 3: DLBENCH defined neural networks

FCN5 is a deep fully connected neural network to classify the MNIST[7] database. It is composed of three layers of 2048, 4096 and 1024 neurons with sigmoid as activation function, which ends with a linear regression to get 10 classes. For convolutional layers, this benchmarks implements a reduced version adapted to CIFAR10[8] images. AlexNet-R[9] includes three convolutional layers (of 3x32, 32x32 and 32x64 widths and heights with stride 5) and a single fully connected with 10 outputs, plus padding, pool, and normalization layers. In the case of ResNet[10], a ResNet-56 is used.

LSTM model is a recurrent network with 2 layers of 256 hidden units each and a sample length of 32

---

[4] http://dlbench.comp.hkbu.edu.hk/

elements. It has been designed to train the Penn Tree Bank (PTB)[11,12] dataset from Tomas Mikolov[5].

The benchmark has been modified to:

a. Accept the path where the input data is available as a parameter of one experiment. Distributed benchmark uses always the HOME directory. This modification permits to study the influence of the file system in the performance of the model.

b. Select the resources to use through the *slurm srun* command, so the available resources for each execution are correctly assigned. Currently, only environment variables are used to control resources by DLBENCH.

c. Select the number of cores for each execution, to allow researchers to make a CPU-only scalability test in a single run or evaluate the influence of the number of cores in a GPU-execution.

d. The caffe-Intel version has been added to the models, to study this framework and the difference with the Berkeley Caffe distribution.

Figure 2 shows a small example of the new format of the input file.

## 3 Methodology

Three kind of experiments have been carried on:

a. Scalability experiments. For each analysed framework, an input experiment configuration was defined, including FCN5, AlexNet and LSTM (when available) networks. For each network, the number of processors or/and the number of GPUs to use was selected. Each execution was performed three times, measuring the execution time as the difference of the dates provided by the Linux OS with the command *date*. The represented times in this report are the mean value of these three executions. Each framework was executed in a single job within a single node in exclusive, having access to all the 4 GPUs and cores. As explained, the resources were limited later using the capabilities of *slurm srun* command for each individual execution.

b. Storage analysis. In this case, two configurations were executed in a single job. One using the LUSTRE filesystem and another accessing the datasets in the EMC cabinet. On both cases, the number of cores or GPUs were changed following the same method as in the previous item.

c. One experiment was executed to study the influence of the batch size. In this case, the batch size and resources were changed during the experiment.

Only FCN5, AlexNet-R and LSTM were executed using Caffe, Caffe-Intel and Tesorflow configurations. For each network, the common parameters among executions have been those in Table 4.

---

[5] http://www.fit.vutbr.cz/~imikolov/rnnlm/simple-examples.tgz

| Network | Learning Rate | Number of epochs | Epoch size |
|---|---|---|---|
| FCN5 | 0.05 | 20 | 60000 |
| AlexNet-R | 0.01 | 20 | 60000 |
| LSTM | 0.1 | 20 | -1 |

Table 4: Common parameters among experiments

```
flag:          testcafee           #Flag of current experiment
tools:         caffe   #Tools to benchmark
experiments: #<network type>; <network name>;  <device id>; <gpu count>; <cpu count>; <batch size>;  <number of epochs>;  <epoch size>; <learning rate>
{
         fc;           fcn5;          0;        1;       1;         1024;          20;
60000;          0.05
}
host_file:     None                #Path to host file or None
cpu_name:      E5-2630v3           #CPU model
device_name:   K80                 #GPU model
cuda:          None                #CUDA version
cudnn:         None                #CUDNN version
cuda_driver:   None                #CUDA driver version
data_dir:      /mnt/lustre/scratch/home/cesga/agomez/dlbench   #data directory
```

Figure 2: Example of an input file for DLBENCH experiments. cuda, cudann, cuda_driver, cpu_name, flag, and host_file entries are informative. Other parameters must be selected.

# 4 Results

## 4.1 Scalability experiments

### Tensorflow (CPU-Only)

Tensorflow 1.2.1 at CESGA has a version which can execute only on CPUs, without GPU support to optimize it and speedup the starting phase. Table 5 shows the elapsed time for the three available models when executed for different number of cores and with a batch size of 512. It shows that the elapsed time increases considerably when the complexity of the model increases as well.

| # cores | FCN5 | AlexNet-R | LSTM |
|---|---|---|---|
| 1 | 1871 | 4593 | 19691 |
| 2 | 1044 | 2736 | 10361 |
| 4 | 564 | 1559 | 5431 |
| 6 | 394 | 1189 | 3715 |
| 12 | 224 | 830 | 2090 |
| 24 | 186 | 768 | 1440 |

Table 5: Elapsed time (s) for Tensorflow 1.2.1 CPU Only

Figure 3, Figure 4, and Figure 5 show the SpeedUp and Efficiency for the three models. Only FCN5 and LSTM models achieve a large efficiency when the number of cores increases. However, AlexNet-R presents a poor scalability with the number of cores for this problem. In any case, use more than 12 cores seems to be unnecessary for this batch size and framework.

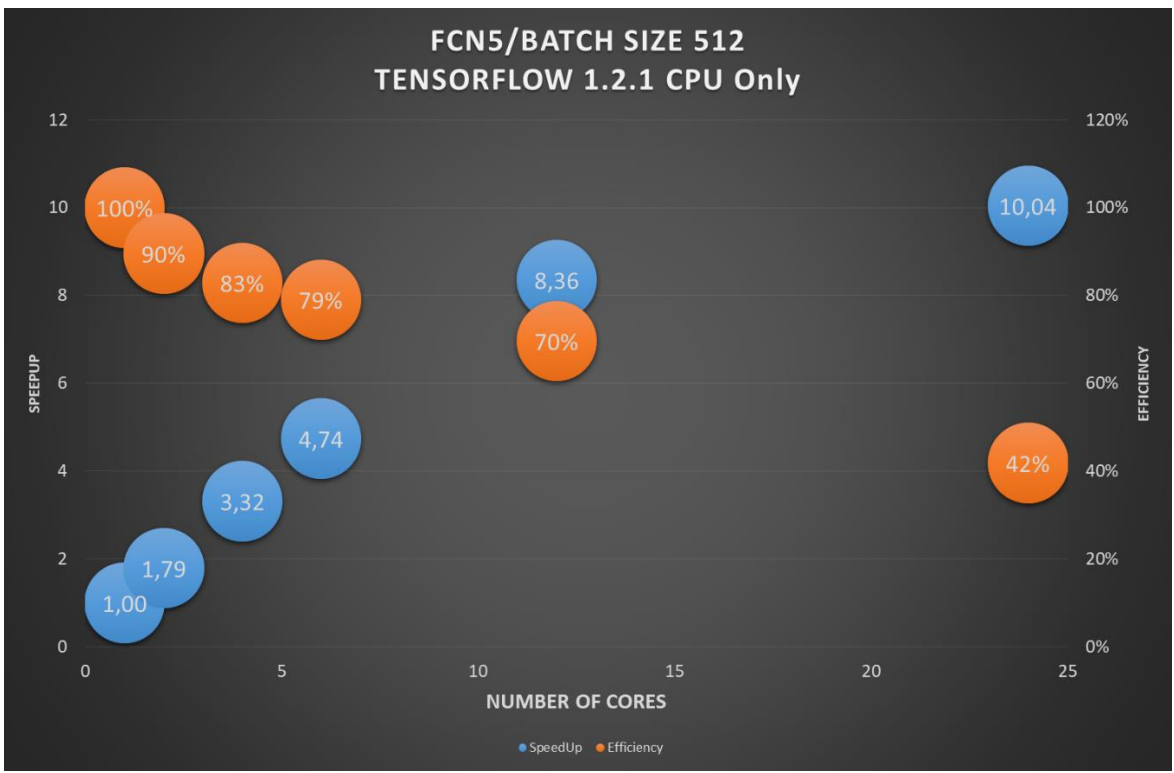

**Figure 3: Tensorflow CPU Only SpeedUp and Efficiency for FCN5 network**

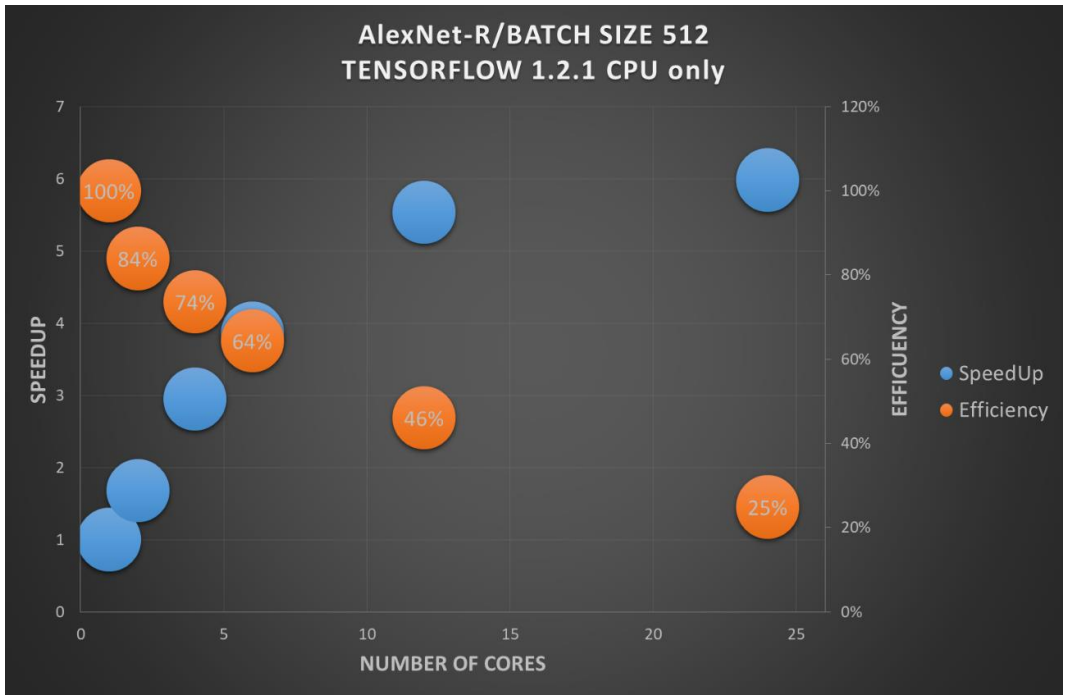

Figure 4: Tensorflow CPU only SpeedUp and Efficiency for AlexNet-R

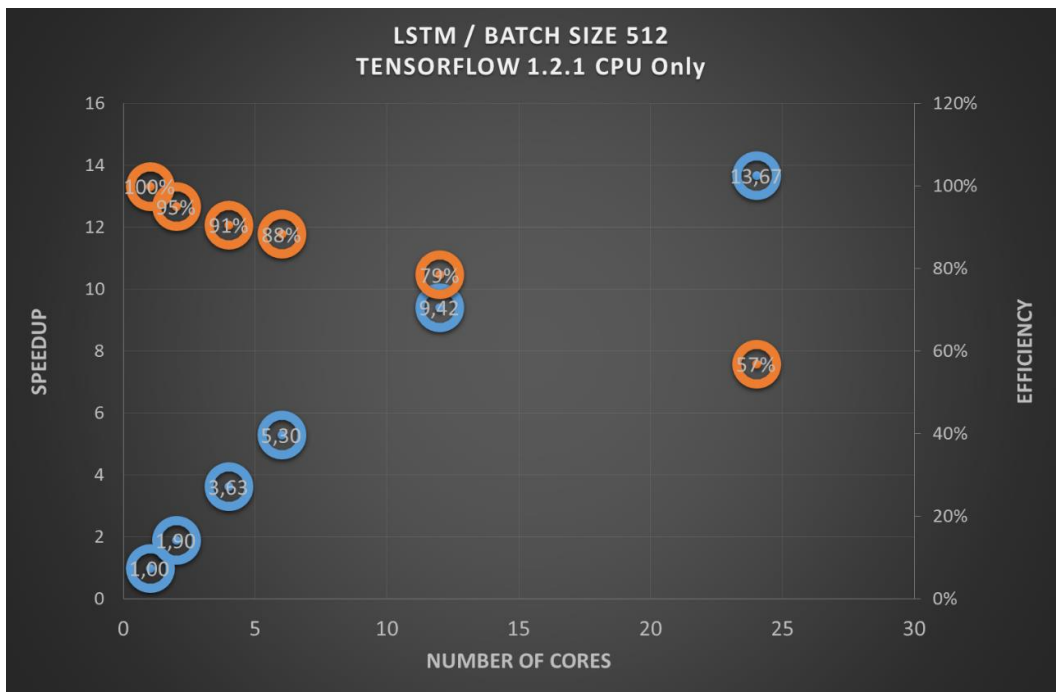

Figure 5: Tensorflow CPU only LSTM SpeepUp and Efficiency

## Caffe

There are two networks available for Caffe: FCN5 an Alexnet-R. The scalability experiments were carried on independently for each network using a single node with two NVIDIA K80 GPUs, that Finis Terrae II detects as four GPUs in terms of resources, numbered as 0,1,2,3. GPU 0 and 1 belongs to the same K80 board, as said before.

Figure 6 shows the execution time in Elapsed Time versus the allocated resources. There exists a small scalability when the number of cores increases from 1 to 24 (around 4 for the maximum number of cores). However, as expected for this framework, the elapsed time decreases a lot when one GPU is used (close to 22 times faster than a single CPU). But, unfortunately, for this problem, increasing the number of GPUs does not help to decrease the elapsed time (although, at least, does not increase). When the achieved accuracy (for 40 Epochs with the same learning rate) is analysed, there is a significant decrease when additional GPUs are included, as it is shown in Figure 7. This effect is well-known and it is due to the increase of the real batch size for step which has not been compensated with an increase of the learning rate.

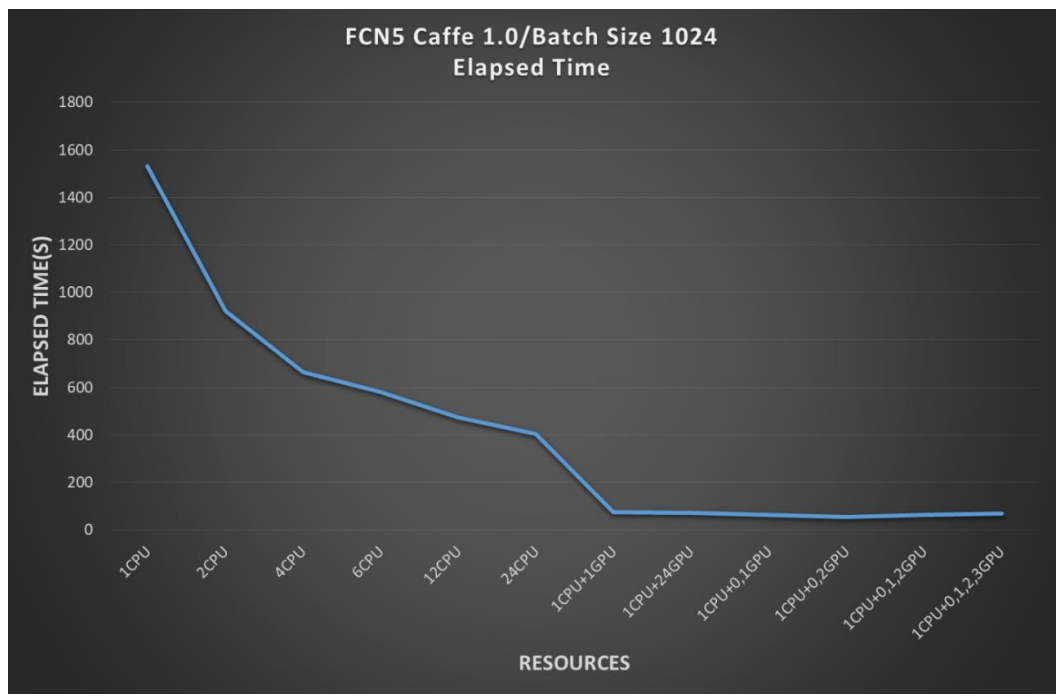

Figure 6: FCN5 execution time for Caffe with batch size of 1024

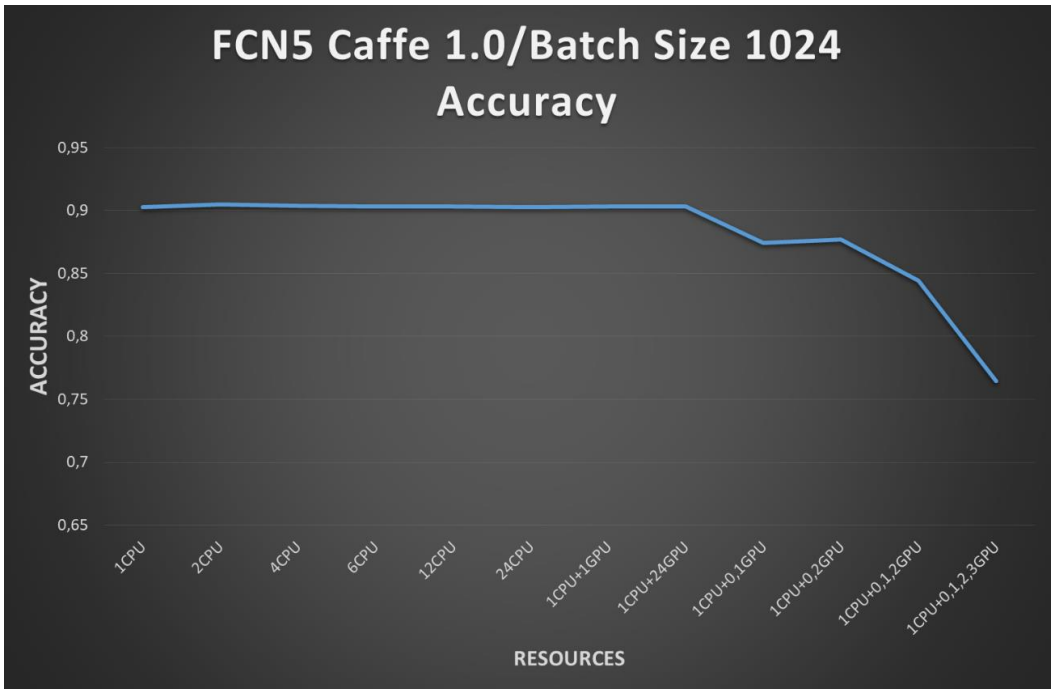

Figure 7: FCN5 Caffe Accuracy vs Resources

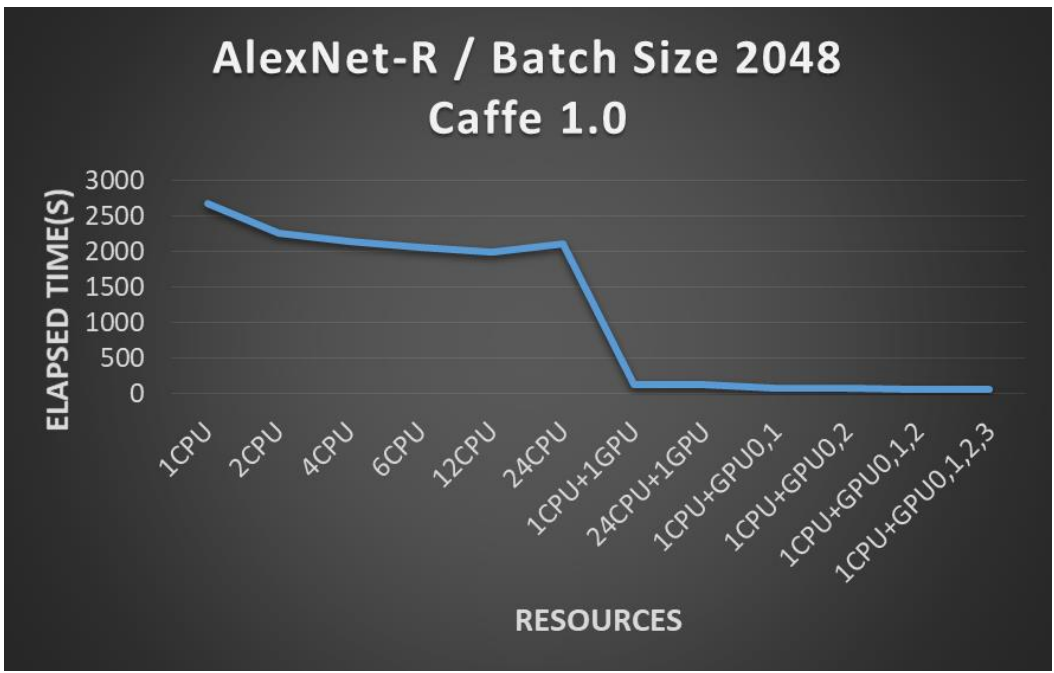

Figure 8: Caffe AlexNet-R Elapsed Time vs resources

Figure 8 shows the performance of Caffe 1.0 for the Alexnet-R neural network. The elapsed time decreases slightly when the number of cores increases, but the speedup in this case is very poor. However, as expected, using one GPU benefits this training problem, reaching a speed-up of 20.5. But, in this case, adding a second GPU decreases the elapsed time with a speed-up of 1.63 when the usage of 2 GPUs is compared with the time for 1 GPU. Adding more GPUs does not significantly help to decrease the time (Figure 9). Figure 10 shows that 1 K80 GPU can process a single image in 127 microseconds.

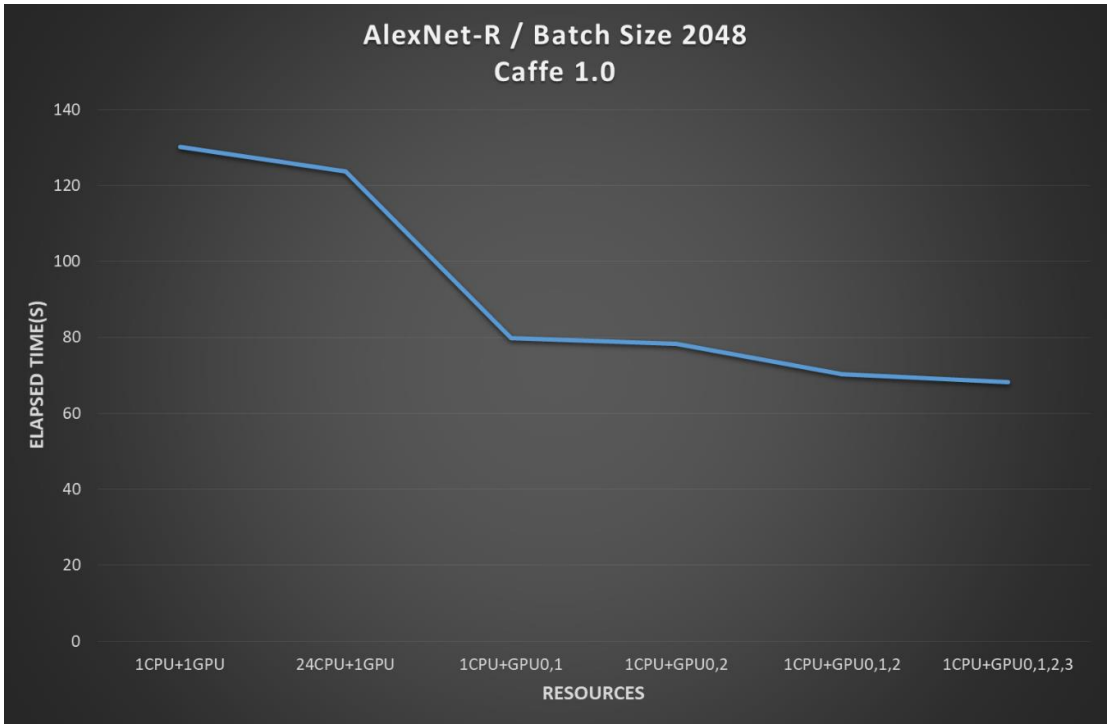

Figure 9: AlexNet-R Caffe 1.0 speed-up vs number of GPUs

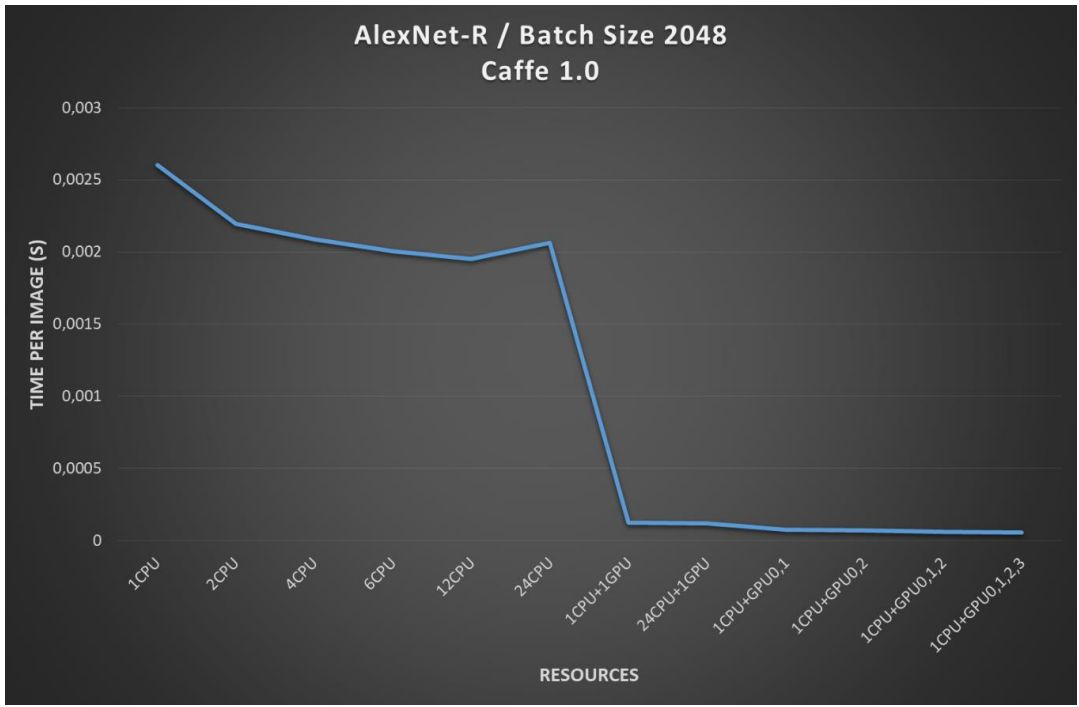

Figure 10: AlexNet-R Caffe Time per Image (s)

## Caffe-intel

Figure 11 shows the elapsed time for FCN5 case for Caffe and Caffe-Intel. Caffe-Intel has a better scalability than Caffe for this case, but Caffe still has a better performance when the GPUs are used. But the differences maybe do not worth the waiting time to access one GPU on Finis Terrae II. However, when the network is changed to AlexNet-R (Figure 12), Caffe-Intel shows a great improvement in the elapsed time. In fact, 24 cores are almost equivalent to one GPU used by Caffe (remember that Caffe Intel has no GPU version).

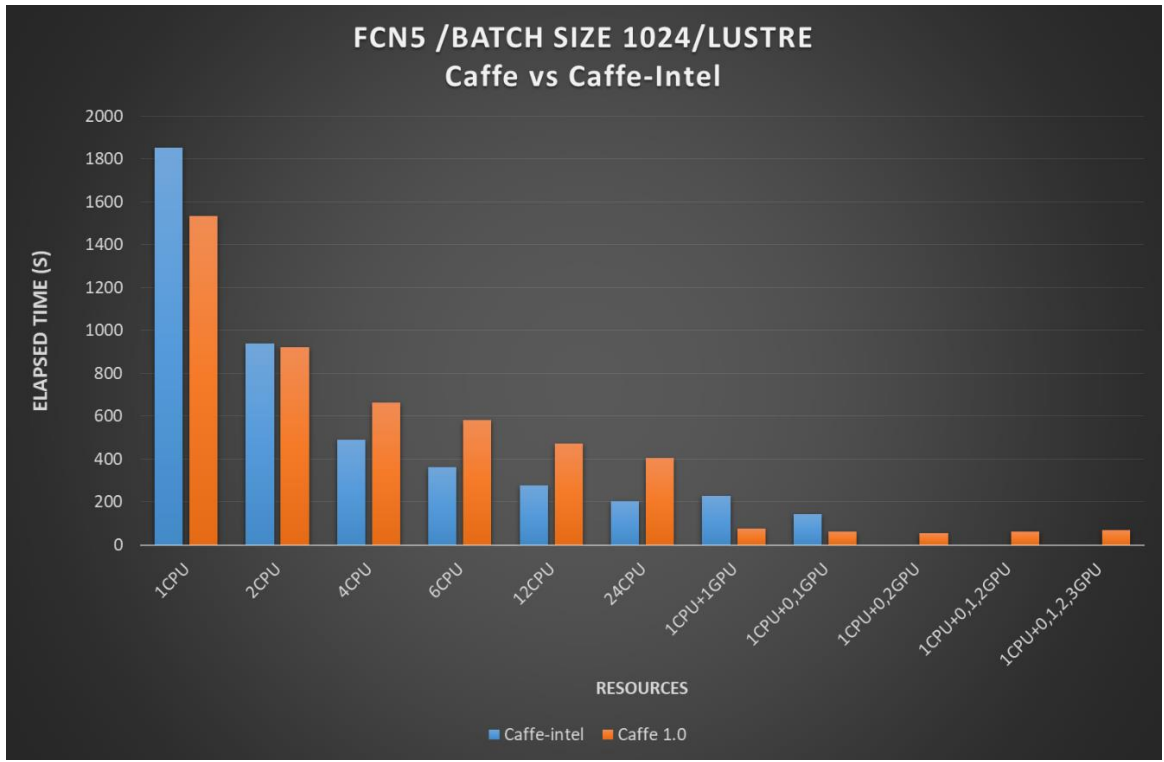

**Figure 11: Caffe vs Caffe-intel for FCN5 elapsed time**

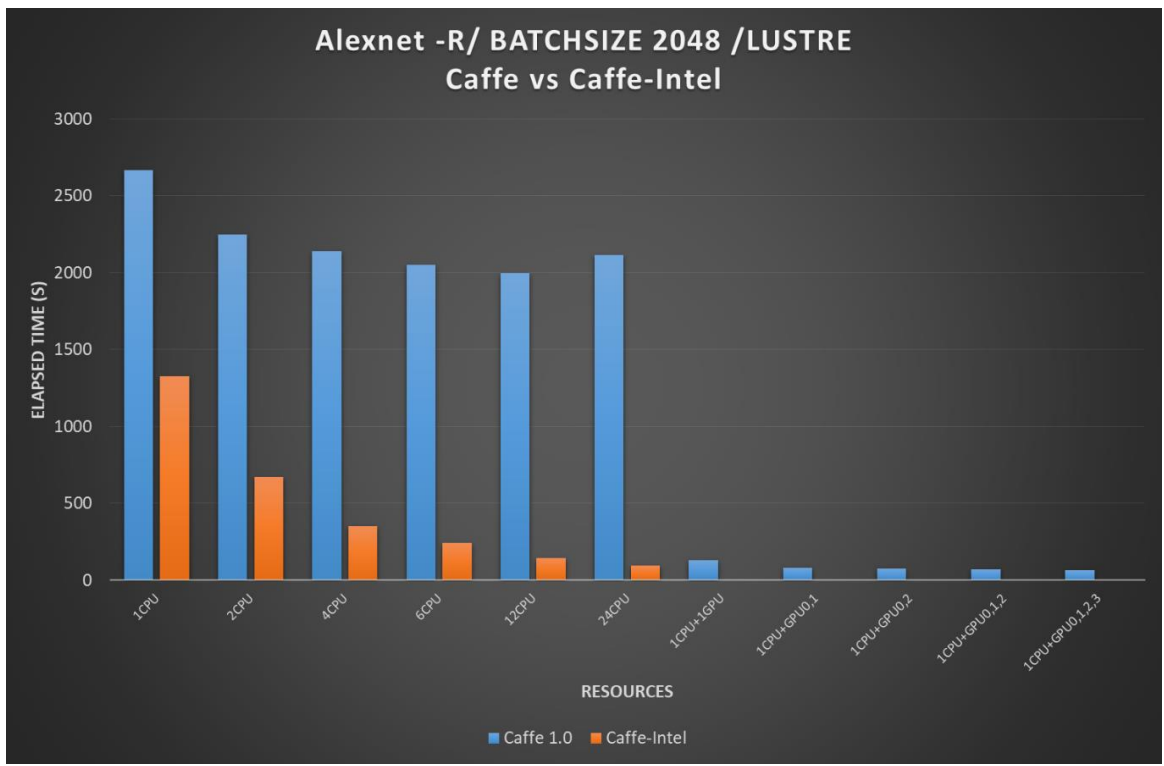

**Figure 12: Caffe vs Caffe-Intel for Alexnet-R**

Certified entity ISO 9001        Page 18 of 24

## 4.2 Storage experiment

This experiment was executed with Caffe-Intel, changing the placement of the input data. Figure 13 demonstrates that there are some small differences when the number of cores increases depending on the file system. Because EMC can manage better small files (as the dataset used by Alexnet-R network), it presents a slightly better speedup when the number of cores increases, showing that can be important to feed correctly the CPUs to gain improvements when enough resources are available.

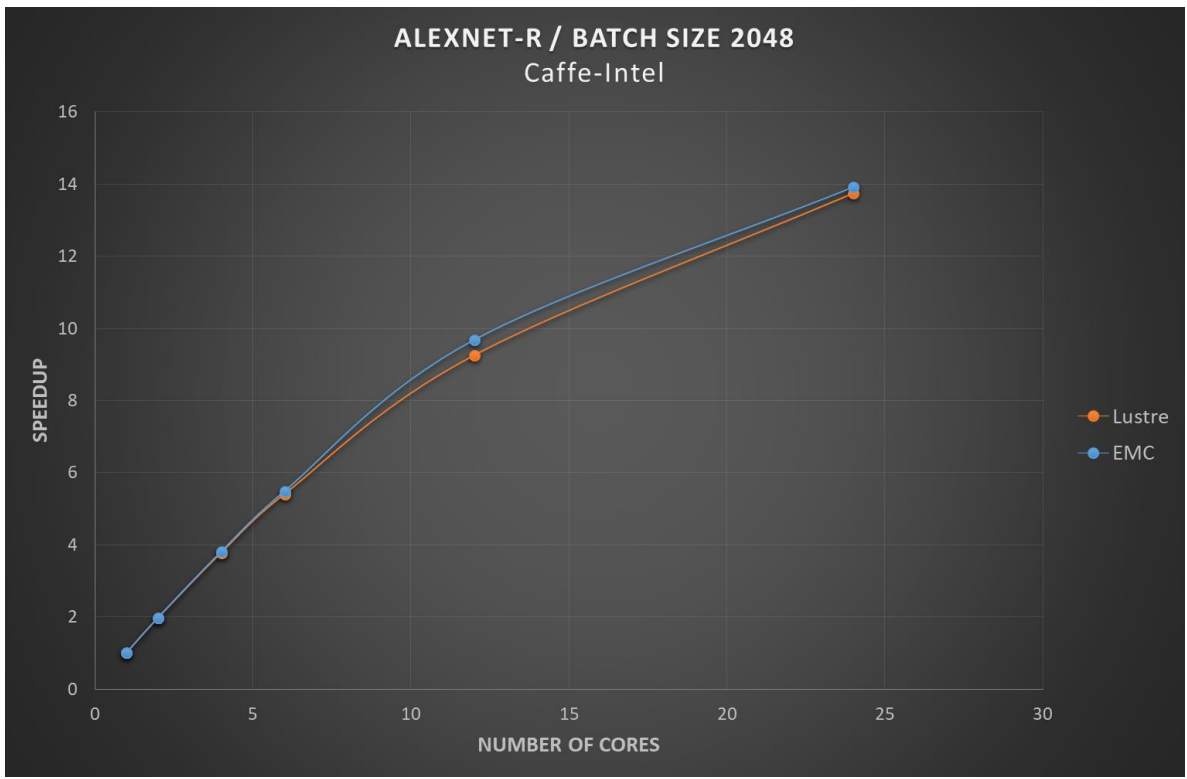

Figure 13: Caffe-Intel Alexnet SpeedUp

## 4.3 Influence of the batch size

In this case, an experiment to evaluate the performance of one model was executed using Tensorflow compiled with GPU support to know better the effects of using different values for the batch size and, additionally, several GPUs. Due to the large number of cases to execute, only a single case per configuration has been performed.

Figure 14 shows the effect of increasing the batch size on the elapsed time and accuracy for a single CPU. As expected, an important decrease of the accuracy is observed because there is no change of the learning rate with the increase of the batch size. So, to use a larger batch size is recommended to adjust accordingly the learning rate to have results with similar quality. Additionally, the elapsed time decreases initially when the batch size increases, but when it is larger than 512, using a single core does not means to have results earlier. Similar situation is observed when a single GPU is used (Figure 15).

Figure 16 shows the speedup of this fully connected training when the number of resources increases, taking also into account the batch size. In this case, increasing the number of cores at the same time that the batch size helps to have a better speedup (up to 12 with 24 cores for batch size of 1024). If instead of using cores, GPUs are used, the speedup versus a single core improves significantly, being around to 24 for a single GPU and 46 when 4 NVIDIA K80 are uses. If the speedup is measured against a single GPU (Figure 17), there is also an improvement when the resources are increased. Using two GPUs inside a single NVIDIA K80 board is slightly worse than using them in different sockets but the difference is really small. Adding the third and fourth GPU still produces some gains, but maybe does not worth.

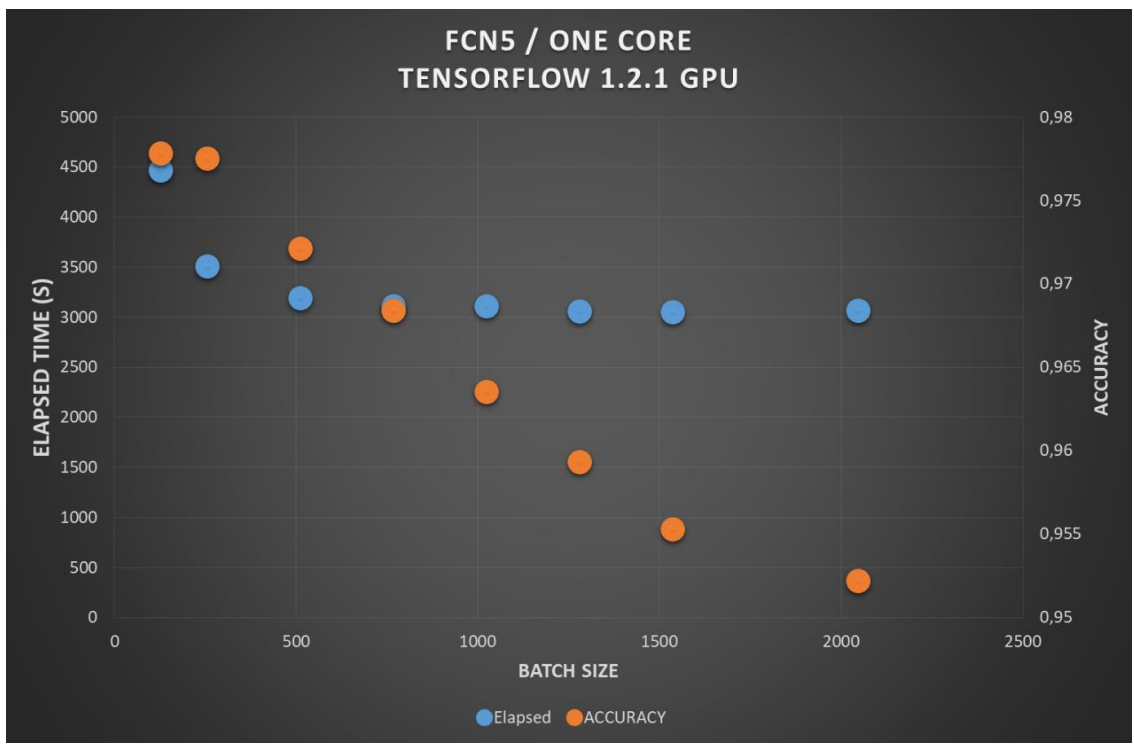

Figure 14: Tensorflow effect of batch size using one core

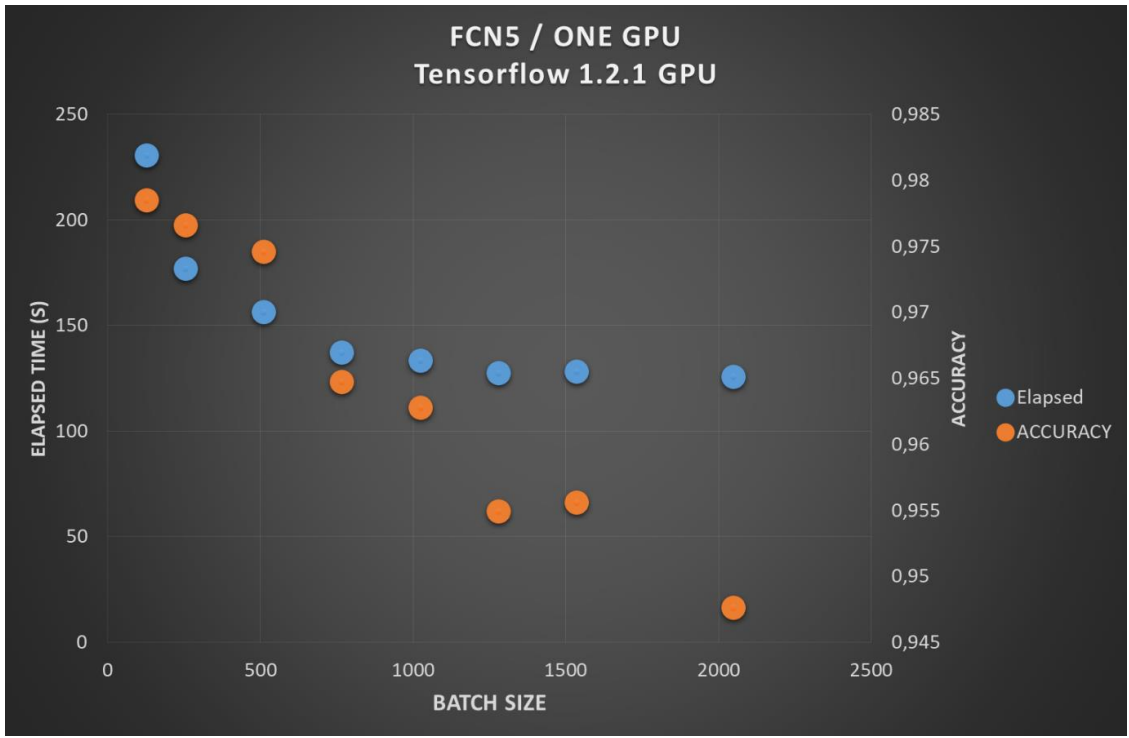

Figure 15: Tensorflow GPU vs Batch size

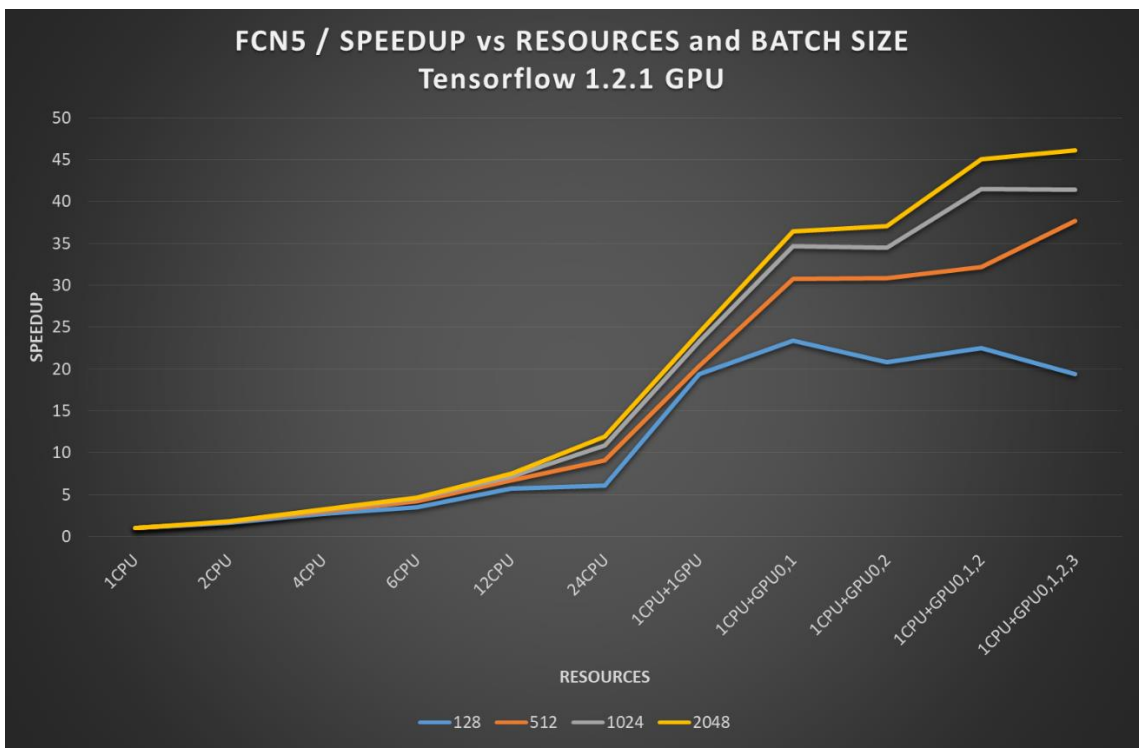

Figure 16: Tensorflow GPU speed-up

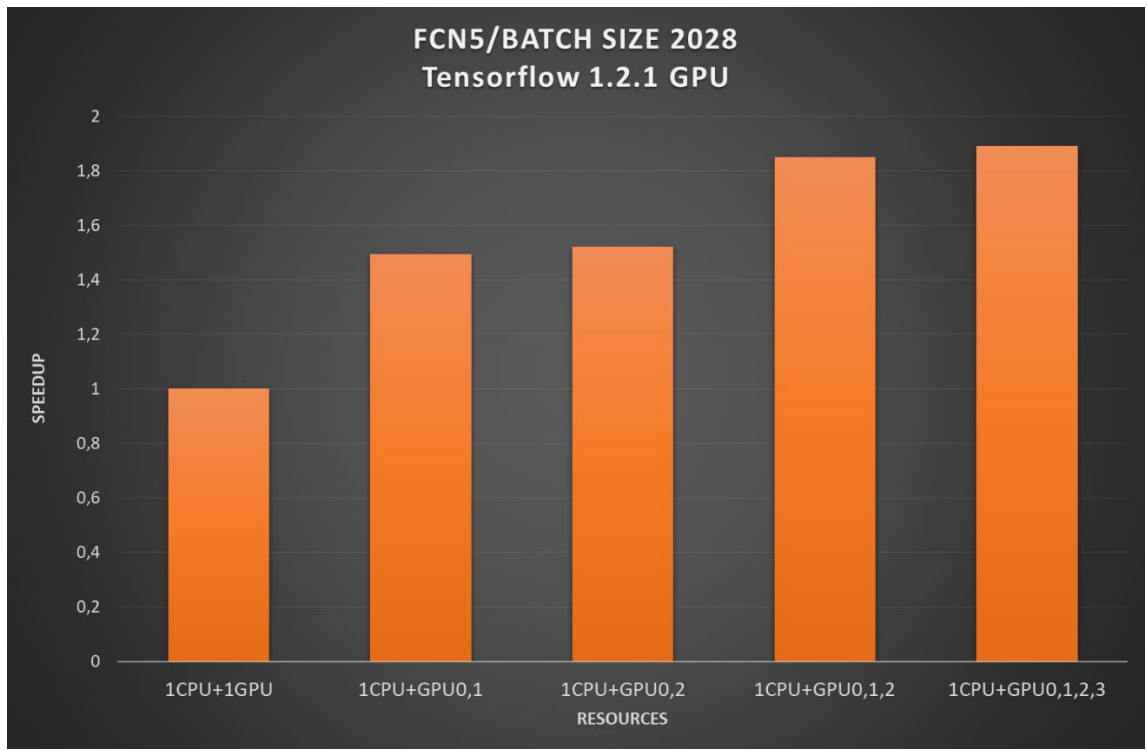

**Figure 17: Tensorflow GPU speedup**

## 5 Conclusions

This initial evaluation of the Machine Learning frameworks on Finis Terrae II supercomputer has shown that:

a. There is a great improvement in the performance of Caffe-Intel version regarding Caffe distribution when only CPU is used. For the evaluated models, Caffe-Intel shows a good scalability when the number of cores increases inside the node, because only threaded parallelism is available. Using Caffe-Intel with 24 cores is almost equivalent to use a single NVIDIA K80 GPU for both analysed models (FCN5 and AlexNet-R). In fact, a single core of Caffe-Intel performs better than 24 cores with Caffe-1.0 for this last network. So, the usage of Caffe 1.0 in multicore machines is not recommended if you can use Caffe-Intel for these convolutional networks.

b. When there is enough computing capacity, a good access to the input data should be guaranteed.

c. During training, a good selection of the batch size is important, because final results could depend on it. An adjustment of the learning rate is usually required when the global batch size is increased. Clearly, increasing the bath size helps to improve the speedup when more resources are added, as has been shown with Tensorflow.

This work has shown the performance of the ML frameworks for different kind of networks and can be used as reference in the selection of the initial resources, data placement or framework when a new Machine Learning project is started. However, this work has no studied the capabilities and performance of these frameworks when several nodes are used to train a single network. Other future benchmarks will address the usage of distributed training with different frameworks as Tensorflow (which uses gRPC) or CNTK (which is based on MPI).

**Acknowledgements**

**References:**

[1] Shi, S., Wang, Q., Xu, P., & Chu, X. (2016). Benchmarking State-of-the-Art Deep Learning Software Tools. arXiv:1608.07249

[2] Jia, Y., Shelhamer, E., Donahue, J., Karayev, S., Long, J., Girshick, R., Darrell, T. (2014). Caffe: Convolutional Architecture for Fast Feature Embedding. arXiv:1408.5093

[3] Abadi, M., et al. (2015): TensorFlow: Large-scale machine learning on heterogeneous systems. Software available from tensorflow.org.

[4] Seide, F., & Agarwal, A. (2016). CNTK: Microsoft's Open-Source Deep-Learning Toolkit. In Proceedings of the 22nd ACM SIGKDD International Conference on Knowledge Discovery and Data Mining - KDD '16 (pp. 2135–2135). New York, New York, USA: ACM Press.

[5] Collobert, R., Kavukcuoglu, K., & Farabet, C. (2011). Torch7: A matlab-like environment for machine learning. BigLearn, NIPS Workshop, 1–6.

[6] Chen, T., Li, M., Li, Y., Lin, M., Wang, N., Wang, M., Zhang, Z. (2015). MXNet: A Flexible and Efficient Machine Learning Library for Heterogeneous Distributed Systems. arXiv:1512.01274

[7] LeCun et al. (1999): The MNIST Dataset of Handwritten Digits (Images)

[8] Krizhevsky, A. (2009): Learning Multiple Layers of Features from Tiny Images

[9] Krizhevsky, A., Sutskever, I., & Hinton, G. E. (2012). ImageNet Classification with Deep Convolutional Neural Networks. Advances In Neural Information Processing Systems, 1–9.

[10] He, K., Zhang, X., Ren, S., & Sun, J. (2015). Deep Residual Learning for Image Recognition. arXiv: 1512.03385

[]Zaremba, W., Sutskever, I., & Vinyals, O. (2014). Recurrent Neural Network Regularization. arXiv:1409.2329

[4] Marcus, M. P., Marcinkiewicz, M. A., & Santorini, B. (1993). Building a large annotated corpus of English: The Penn Treebank. *Computational linguistics*, *19*(2), 313-330.